\documentclass[conference]{IEEEtran}
\IEEEoverridecommandlockouts
\usepackage{subfigure}
\usepackage{multirow}
\usepackage{booktabs}
\usepackage{cite}
\usepackage{amsmath,amssymb,amsfonts}
\usepackage{algorithmic}
\usepackage{graphicx}
\usepackage{textcomp}
\usepackage{xcolor}

\usepackage{wrapfig}
\usepackage{float}
\usepackage{comment}

\usepackage[hyphens]{url}
\usepackage[hidelinks]{hyperref}
\hypersetup{breaklinks=true}
\usepackage{cite}

\def\BibTeX{{\rm B\kern-.05em{\sc i\kern-.025em b}\kern-.08em
    T\kern-.1667em\lower.7ex\hbox{E}\kern-.125emX}}
\begin{document}

\title{Denoising Induction Motor Sounds Using an Autoencoder \\

}

\makeatletter
\newcommand{\linebreakand}{%
  \end{@IEEEauthorhalign}
  \hfill\mbox{}\par
  \mbox{}\hfill\begin{@IEEEauthorhalign}
}
\makeatother


\author{\IEEEauthorblockN{Anonymous Authors}}
\author{
    \IEEEauthorblockN{Thanh Tran\IEEEauthorrefmark{1}, Sebastian Bader\IEEEauthorrefmark{1}, Jan Lundgren\IEEEauthorrefmark{1}}
    \IEEEauthorblockA{\IEEEauthorrefmark{1} STC Research Centre, Department of Electronics Design, Mid Sweden University
    \\\{thanh.tran, sebastian.bader, jan.lundgren\}@miun.se}}

\maketitle

\begin{abstract}
Denoising is the process of removing noise from sound signals while improving the quality and adequacy of the sound signals. Denoising sound has many applications in speech processing, sound events classification, and machine failure detection systems. This paper describes a method for creating an autoencoder to map noisy machine sounds to clean sounds for denoising purposes. There are several types of noise in sounds, for example, environmental noise and generated frequency-dependent noise from signal processing methods. Noise generated by environmental activities is environmental noise. In the factory, environmental noise can be created by vehicles, drilling, people working or talking in the survey area, wind, and flowing water. Those noises appear as spikes in the sound record. In the scope of this paper, we demonstrate the removal of generated noise with Gaussian distribution and the environmental noise with a specific example of the water sink faucet noise from the induction motor sounds. The proposed method was trained and verified on 49 normal function sounds and 197 horizontal misalignment fault sounds from the Machinery Fault Database (MAFAULDA). 
The mean square error (MSE) was used as the assessment criteria to evaluate the similarity between denoised sounds using the proposed autoencoder and the original sounds in the test set. The MSE is below or equal to 0.14 when denoise both types of noises on 15 testing sounds of the normal function category. The MSE  is below or equal to 0.15 when denoising 60 testing sounds on the horizontal misalignment fault category. The low MSE shows that both the generated Gaussian noise and the environmental noise were almost removed from the original sounds with the proposed trained autoencoder. 
\end{abstract}

\begin{IEEEkeywords}
Autoencoder, Denoise sound.
\end{IEEEkeywords}

\section{Introduction}

In the manufacturing industry, machine failure detection systems are critical for automatically detecting broken components in machines for scheduled maintenance~\cite{8908771, 9128057}. This can minimize the downtime of the machine as well as the maintenance cost. Researchers have recently focused on exploiting emitted sounds to detect machine failures. Noise, on the other hand, is present in the sound of machines recorded directly in factories. White noise or other background noise from the recording environment might be included in the signal (the noise of talking, the water sink faucet sound, the sound of the wind, knocking on the microphone, etc.). If the noise of the sound is removed, the quality of the sound event will increase, resulting in the successful application of machine failure detection systems based on sound. In this paper, a method is described for denoising machine sounds by mapping noisy sounds to clean ones using an autoencoder.


In recent decades, researchers have been working to solve the denoising problem. For many years, there has been a huge growth of many conventional methods to reduce noise in sounds and speech, for example, minimum mean square error (MMSE)~\cite{4518541}, spectral subtraction~\cite{4696866, 9015244, 6689694, 7488382}, Wiener filtering~\cite{5937517, 7009836, 1594205}, time-frequency block threshold~\cite{4490115}, nearest neighbor estimation~\cite{6376717}, Singular Value Decomposition (SVD)~\cite{6217853}, and double-density dual-tree discrete wavelet transform (DDDTWT)~\cite{7455802}.

Although the conventional denoising methods have achieved good performance in image, sound, and speech denoising, they still have some drawbacks. These conventional methods are only reasonably effective for images and speech with a low level of noise. Besides, noise estimation and assumptions about the aggregate statistical models are fundamental to these conventional techniques. Hence, these algorithms frequently underestimate or overestimate noise, which results in either insufficient noise removal (noise audible in the filtered result) or audio distortions caused by excessive noise removal.
Moreover, identifying the gain of a Wiener filter requires knowing the power spectral densities (PSDs) of the noise and the desired signals at a particular frequency. The SVD computation is slow and has a high computational cost. The choice of the proper wavelets when denoising a sound signal using wavelet transform, for example, can be time-consuming~\cite{Wavelet}.
By using deep neural networks (DNNs) to denoise, these problems can be solved. DNN-based methods use paired data of noisy sounds and their corresponding clean sounds to train their denoising model. They outperform conventional denoising filters to denoise images~\cite{NIPS2008_c16a5320} and signals~\cite{8466616, Liu2014ExperimentsOD, Park2017AFC, 8683648, Fakoor2017ReinforcementLT, 9247199}. Specifically, deep learning can effectively remove the issue of Gaussian noise in sounds. In the supervised learning approach, pre-trained DNNs combined with linear support vector machines have been used to learn and predict features of noisy signals~\cite{6473841}.
Convolutional neural networks (CNNs) are attracting considerable interest due to many applications in signal processing, image processing, and computer vision. Besides, CNNs have applications in the field of denoising. The training of CNNs on a large number of noisy images can enhance their ability to adapt to different standard noise, thus allowing for a greater degree of generalization~\cite{ilesanmi_ilesanmi_2021}. Some hybrid methods that are the combination between conventional denoising methods and deep learning methods have been proposed to denoise sounds~\cite{Tu2019TowardsAL, 8547084, nie18_interspeech}.

Denoising autoencoders have been applied to reduce added noise in images~\cite{Fan2019BriefRO, khalaf2018deep, song2020image, 7279746, team}. For denoising images, autoencoders have shown better performance than conventional noise filters because autoencoder can be modified based on the input images,  and hence could be labeled to be data specific. 
For denoising images, images are corrupted by adding random noise, and the autoencoder is trained on the original images and the corresponding noisy images (corrupted images) to produce noise-free images (uncorrupted images). 
Recently, denoising autoencoders have been used successfully to denoise ECG signal~\cite{8693790, XIONG2016194, antczak_2019} and radio signals~\cite{Wang2019ModulationCB, Almazrouei2019ADL, Almazrouei2019UsingAF}.

The noise in sound is frequently assumed to follow a Gaussian distribution in conventional denoising algorithms. Real noise in sound, on the other hand, is more complicated. Machine sounds in factories, for example, are rarely particular to the machine and may include ambient noise such as human-produced sounds or sounds from other machines. The spectral content of background noise and the sound event can sometimes overlap. Conventional denoising approaches may not perform well in these situations because they may remove portions of the sound event that sound like background noise. Due to this, there may be audible distortions as a result of the denoising process. Hence, effective denoising technologies that can eliminate various types of environmental noise (background noise) and improve sound intelligibility and recognition rate but not cause sound distortion are still needed.

The autoencoder's effectiveness in other applications motivates its use in sound signal denoising, particularly for canceling random white noise and various types of acoustic noise in real life. As a result, the goal of this research is to develop, implement, and evaluate an autoencoder for reducing sound noise and thereby improving the quality of sound events. We attempt to denoise sounds as close to their original clean sound as possible by learning the noise distribution in sounds and denoising them. To create noisy sound, two types of noise were added to the original clean sound in this study. The autoencoder is then trained using the original clean sound and the corresponding noisy sound. The autoencoder can recover the noisy sound as close as the original clean sound after training. To our knowledge, this is the first study to evaluate a denoising autoencoder for noise reduction in industrial motor sounds.

The rest of this paper is organized as follows. A brief overview of the dataset is presented in Section II. The proposed method is described in Section III. The experimental result is presented in Section IV. The discussion is written in Section V. The conclusion is drawn in Section VI.

\section{Dataset}
The MAFAULDA dataset~\cite{mafaulda} includes the time series that are acquired from the "SpectraQuest’s Machinery Fault Simulator (MFS) Alignment-Balance-Vibration (ABVT)" using different sensors, including microphones. The data acquisition system includes three industrial IMI sensors (model 601A01 accelerometers), one triaxial accelerometer (model 604B31), Monarch Instrument MT-190 analog tachometer, Shure SM81 microphone, and two National Instruments NI 9234 4 channel analog acquisition modules with a sample rate of 51\kern 0.16667em200 Hz. The dataset is divided into six categories: normal function, imbalance faults, vertical misalignment faults, horizontal misalignment faults, inner bearing faults, and outer bearing faults. Due to the application aspect in detecting and classifying industrial motor faults, this dataset was selected for this research. There are 49 sounds in the normal function category when the machine is working properly. There are 197 sounds in the horizontal misalignment category, divided into four types: 0.5mm misalignment (50 sounds), 1.0mm misalignment (49 sounds), 1.5mm misalignment (49 sounds), and 2.0mm misalignment (49 sounds). For this first investigation, we focused on these two categories (the normal function and the horizontal misalignment faults), but we see no reason that the same approach should not work also within the other classes. 

The sound signals were recorded by Shure SM81 microphones with a sample rate of 50\kern 0.16667em000 Hz and stored in comma-separated values (CSV) files. Before further analysis, the CSV files were converted into waveform audio file format (WAV) files. Each signal was sampled at a rate of 50\kern 0.16667em000 Hz for 5 seconds, resulting in a time series of 250\kern 0.16667em000 samples.

\section{Methodology}
An encoder and a decoder are the two halves of an autoencoder. The encoder compresses the data that it receives. The decoder uses the encoder's latent representation to reconstruct the output. Noise is added to the original sounds to create noisy sounds as input for the denoising autoencoder. To train the denoising model, the noisy sounds and their corresponding clean sounds were fed into an autoencoder. The encoder learns the noisy sounds during the encoding process, while the decoder attempts to generate new sounds that best match the clean input sound.

The first stage is to add noise to the original induction motor sounds. This paper aims to implement a denoising autoencoder with two types of noise: random white Gaussian noise and water sink faucet noise.

The random noise was generated stochastically from a normal Gaussian distribution where \(0\) is the mean and \(1\) is the standard deviation. The length of each random noise was equal to the length of the sound in the dataset. This noise is multiplied with the noise factor of 0.1 and then added to the original sounds to create corresponding noisy sounds. Hence, each sound in the dataset was superimposed with a different random noise.

Environmental noise is usually the pink noise (for example, heartbeats, steady rain, the wind blowing through trees, the sound of leaves rustling, waves crashing on the shore), brown noise (thunder), or blue noise (water hissing out of a faucet). White noise is different from environmental noise, as it has equal power over the entire frequency band (the TV or radio static). High-frequency white noise is sometimes referred to as blue noise.

The water hissing out of a faucet is a common noise in some factories because various industries require large amounts of water to manufacture, process, wash, dilute, cool, or transport products. Harmonic vibration caused by water flowing past the valve seat and seal is the most likely cause of an unusual noise from a faucet. Hence, a denoising autoencoder was trained with a specific type of noise (the sound of the water sink faucet as an example) instead of using the random white noise. The water sink faucet sound is the blue noise. The spectral density (power per hertz) of blue noise is proportional to its frequency. As the frequency increases, the energy and power of the signal increase. 
The water sink faucet noise is 5 seconds in length, the same as the length of the original sound in the dataset (Figure~\ref{fig:sinkfaucet}). 
Figure 1 plots the amplitude envelope of a waveform using a \emph{librosa.display.waveplot} \cite{librosa.display.waveplot} in which the sound signal is down-sampled and the envelope is visualized rather than the signal itself. When y is monophonic, a filled curve is drawn between [-abs(y), abs(y)].

\begin{figure}[hpt!]
    \centering
    \includegraphics[height=4.5cm]{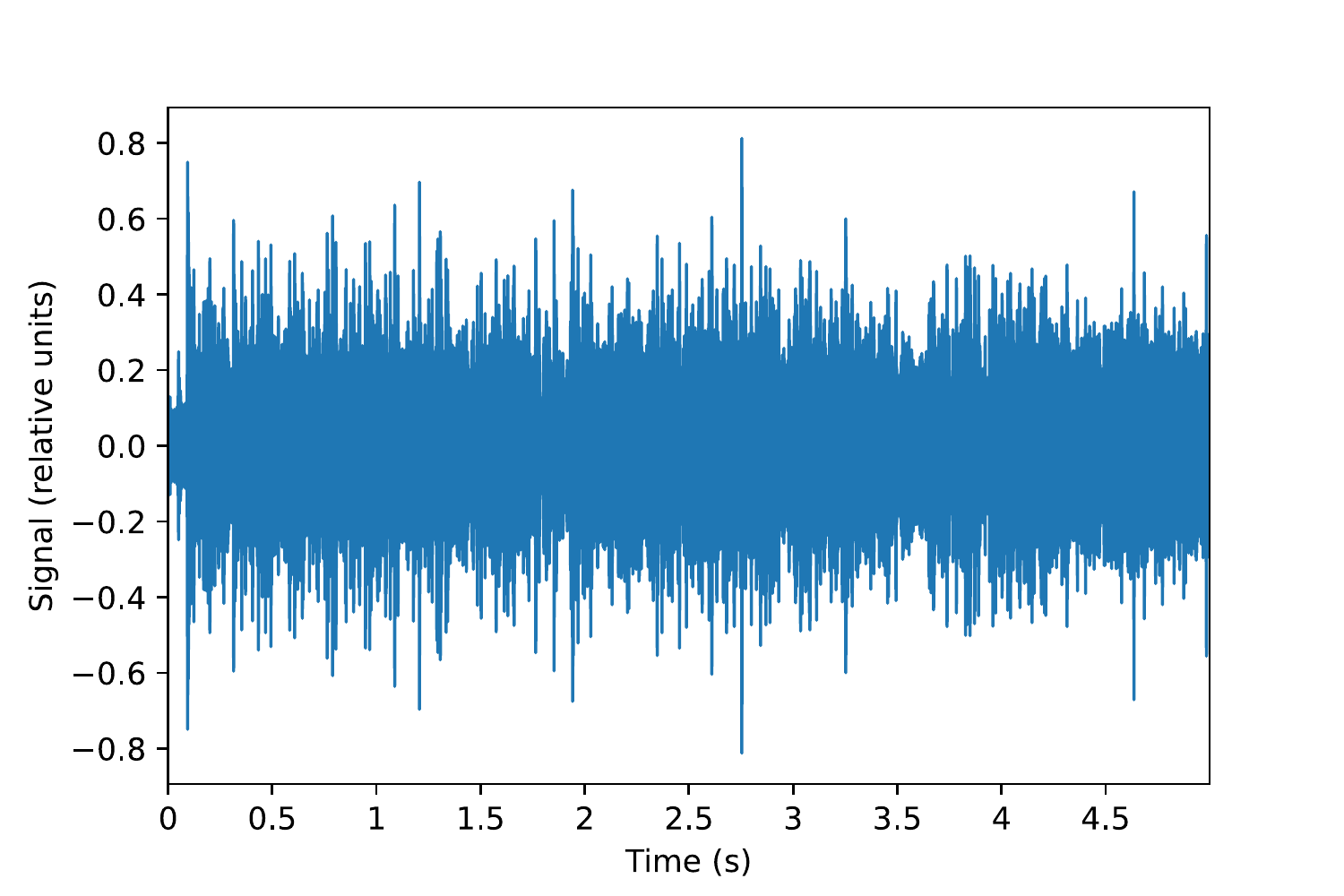}
    \caption{The water sink faucet noise.}
    \label{fig:sinkfaucet}
\end{figure}


The encoder consists of four convolutional 1D layers with a kernel size of 3. The first convolutional 1D layer has a filter size of 128 and a rectifier linear unit (ReLU) as an activation function. The input size of the encoder is (250\kern 0.16667em000, 1), corresponding to 250\kern 0.16667em000 samples in each time series. The second, third, and fourth convolutional 1D layers have the filter size of 32, 16, and 8, respectively. The decoder consists of four convolutional 1D transpose layers and a convolutional 1D layer. The first, second, third, and fourth convolutional 1D transpose layers have the filter size of 8, 16, 32, and 128, respectively. The final convolutional layer turns the number of channels back into one. The batch size is 8. The number of epochs is 10. 
We used 30\% of sounds for testing, 56\% for training, and 14\% for validation. The max norm constraints regularization was utilized with a value of 2.0 so the updates of the network are always bounded.  Using projected gradient descent to enforce the constraint, maximum norm constraints establishes an absolute upper limit on each neuron's weight vector. The binary cross-entropy was utilized as the loss function. Hence, the original sounds and noisy sounds were normalized so that their values are in the range of [0, 1] using this formula:

\begin{equation}
    n_{i} = (y_{i} - min(y))/(max(y) - min(y)),
\end{equation}
where all variables are defined as follows:
\begin{itemize}
    \item $n_{i}$: the normalized value $i_{th}$ in the dataset;
    \item $y_{i}$: the sample value $i_{th}$ in the dataset;
    \item $min(y)$: the minimum value in the dataset;
    \item $max(y)$: the maximum value in the dataset.
\end{itemize}

The autoencoder was compiled using the Adam optimizer~\cite{kingma2017adam} and binary cross-entropy loss function. Adam is an optimization algorithm for updating network weights iteratively in response to the training data. Adam provides an optimization solution to noisy problems that handles sparse gradients by combining the Adaptive Gradient Algorithm and Root Mean Square Propagation. The binary cross-entropy loss function was utilized instead of root mean squared error because the last layer of the proposed autoencoder uses a sigmoid activation function. The binary cross-entropy loss function ($bce$) is calculated as follow: 
\[bce = -\frac{1}{N}\sum_{i=1}^{N}(y_{i}\cdot\log \hat{y_{i}}+(1-y_{i})\cdot\log (1-\hat{y_{i}})),\]
where $N$ is the number of sounds, $\hat{y_{i}}$ is the output value of the model and $y_{i}$ is the corresponding target value.

\begin{figure*}[hpt!]
    \centering
    \includegraphics[width=0.85\linewidth]{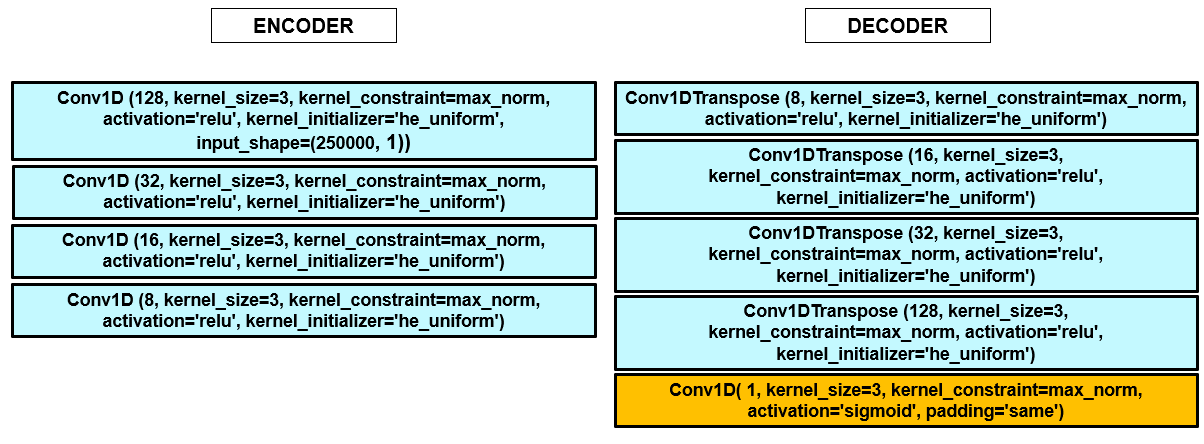}
    \caption{The autoencoder architecture.}
    \label{fig:autoencoder}
\end{figure*}


\section{Experimental results}
The autoencoder was implemented using Keras and TensorFlow, Librosa for audio processing, Matplotlib, Numpy, and the Python math library. For the conversion of audio files into readable arrays, the input sampling rate was 50\kern 0.16667em000 Hz. Each specific category of the dataset was trained with the same autoencoder to denoise the sound. During the decoding process, the autoencoder computes and minimizes the difference between the original noise-free sound (pure sound) and the reconstructed sound (denoised sound).

The mean square error (MSE) was used as the quantitative evaluation criteria for the model. MSE is the average of the square of the difference between the original and denoised sounds generated by the autoencoder. Low MSE indicates more accurate forecasting by the autoencoder. MSE is calculated as follows:
\begin{equation}
    MSE = \frac{1}{n}\sum_{i=1}^{n}(y_{i} - \hat{y_{i}})^2,
\end{equation}
where $n$ is the total number of sounds in the test set, $\hat{y_{i}}$ is the output value of the model, and $y_{i}$ is the corresponding target value.

\subsection{The normal function category}
Seventy percent of the sounds in the normal function category were used for training and validation (34 sounds) and thirty percent for testing (15 sounds). The training and validation loss of the autoencoder when denoise random Gaussian noise and water sink faucet noise from sounds in the normal function category is visualized in Figure~\ref{fig:loss}a and Figure~\ref{fig:loss}b, respectively. Overfitting did not occur during the training process. The autoencoder converges to an acceptable validation loss of 0.66 after 6 epochs. There was little improvement in the autoencoder's performance after 6 epochs. To find out how well the model works, noisy sounds from the test set, which are sounds the autoencoder has never seen before, were fed into the trained autoencoder to denoise. 

\subsubsection{Denoise random Gaussian noise}
\begin{figure*}[hpt!]
    \centering
    \subfigure[]{\includegraphics[width=0.4\linewidth]{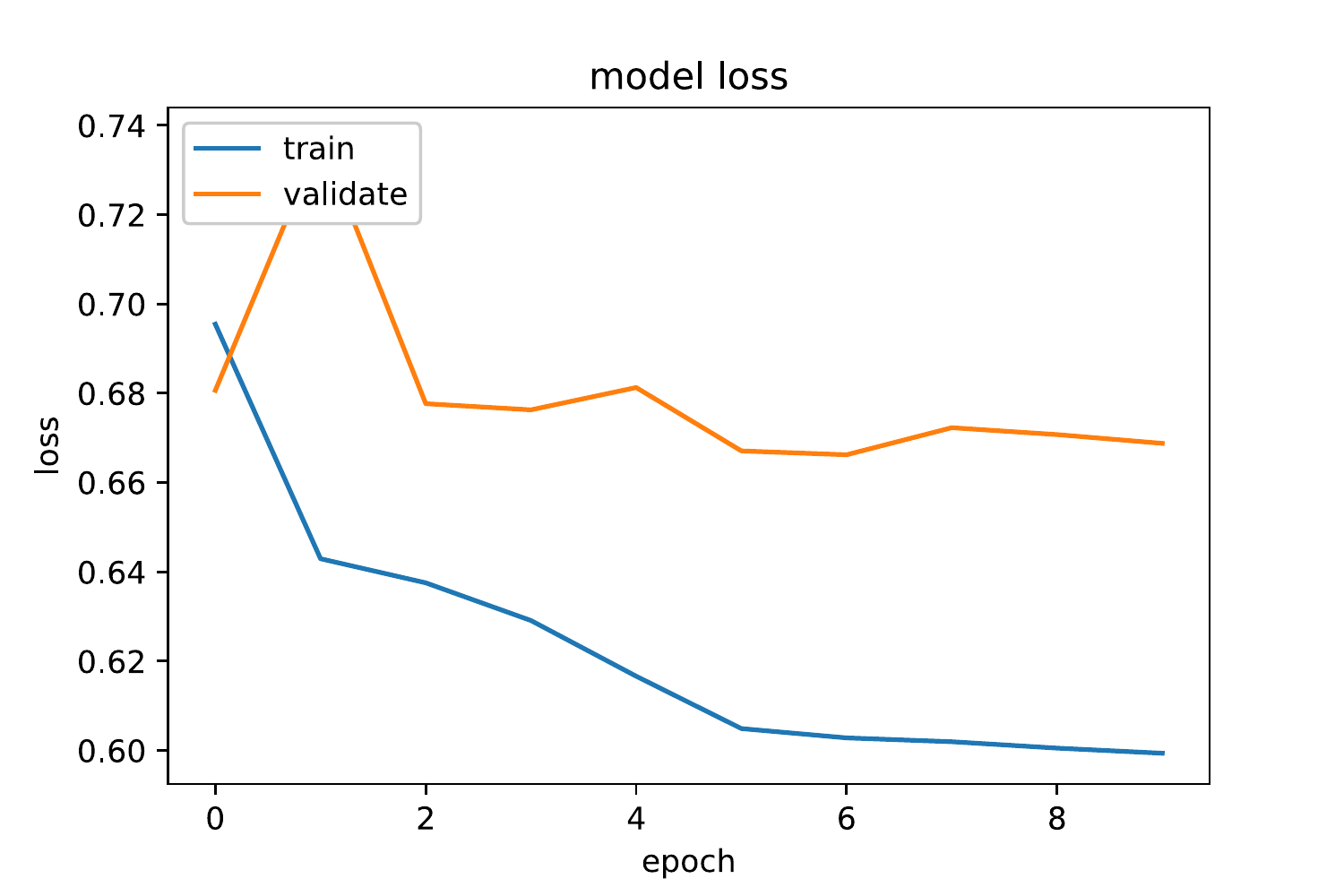}} 
    \subfigure[]{\includegraphics[width=0.4\linewidth]{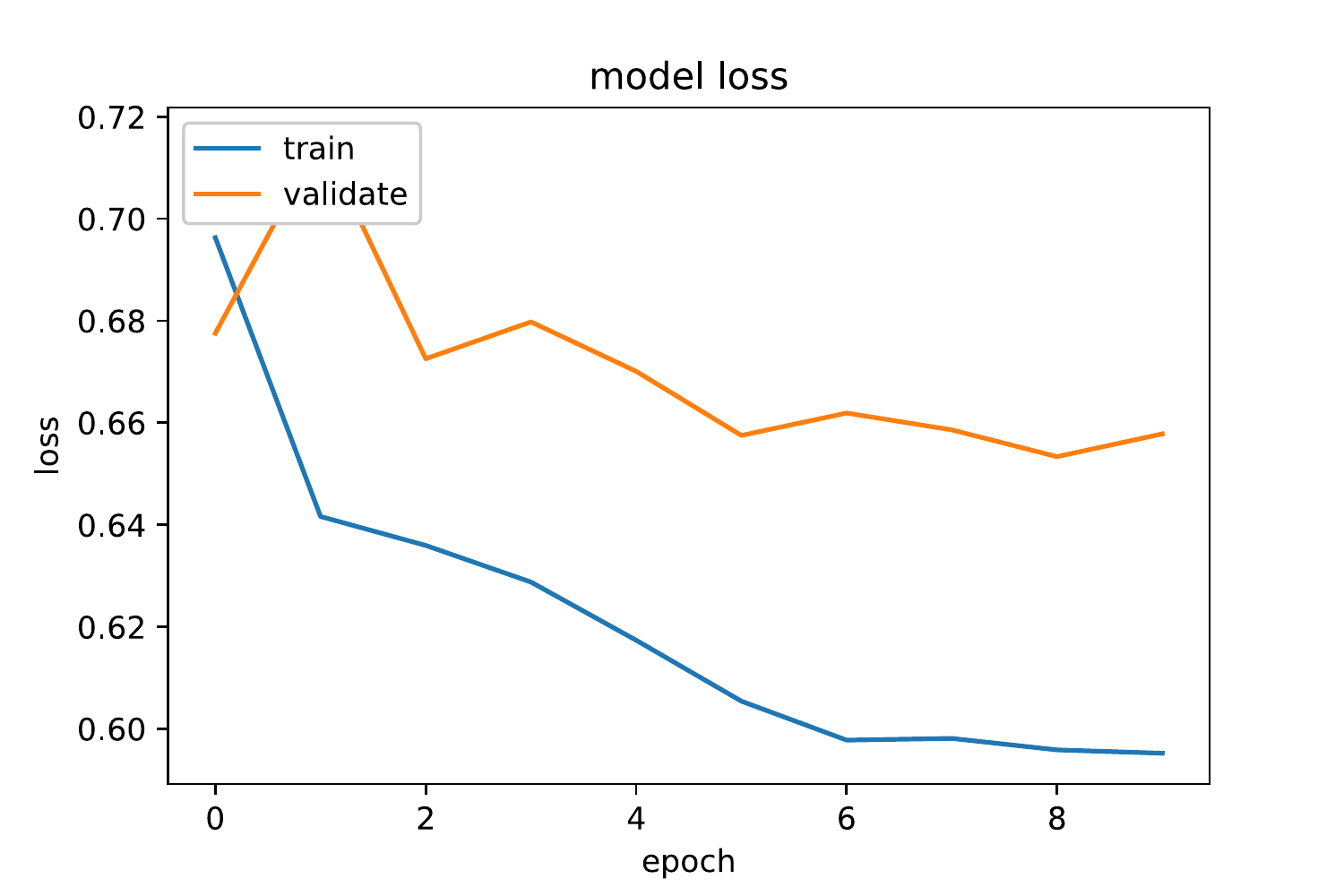}} 
    \caption{Training and validation loss for removing: (a) the random Gaussian noise (b) the water sink faucet noise on the normal function category.}
    \label{fig:loss}
\end{figure*}

The waveform of the original noise-free sounds, noisy sounds, and denoised sounds are visualized in Figure~\ref{fig:result_randomnoise}a. The corresponding STFT power spectrum is visualized in Figure~\ref{fig:result_randomnoise}b. The STFT power spectrum of the reconstruction sounds (denoised sounds) from the trained autoencoder is visually much more similar to the STFT power spectrum of the original sounds in the dataset. 

\begin{figure*}[hpt]
    \subfigure[]{\includegraphics[width=0.9\textwidth]{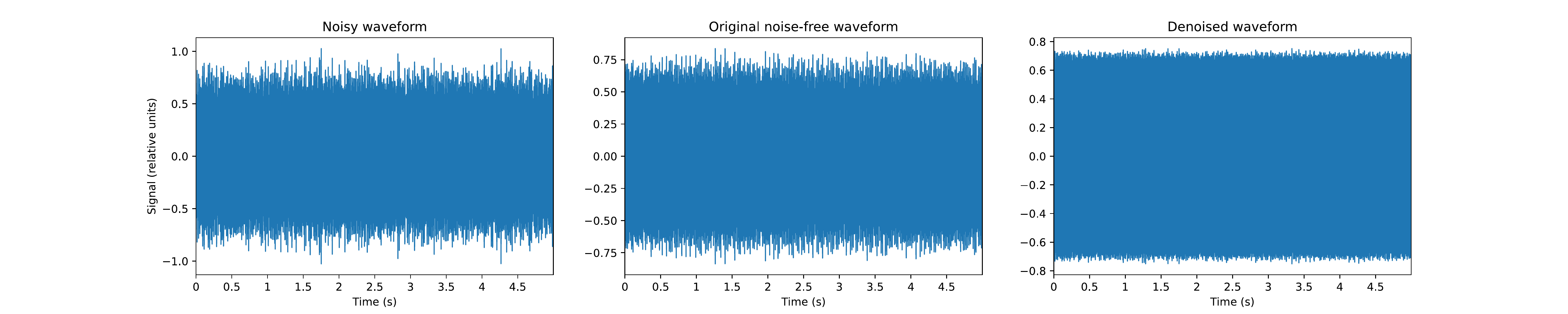}} 
    \subfigure[]{\includegraphics[width=\textwidth]{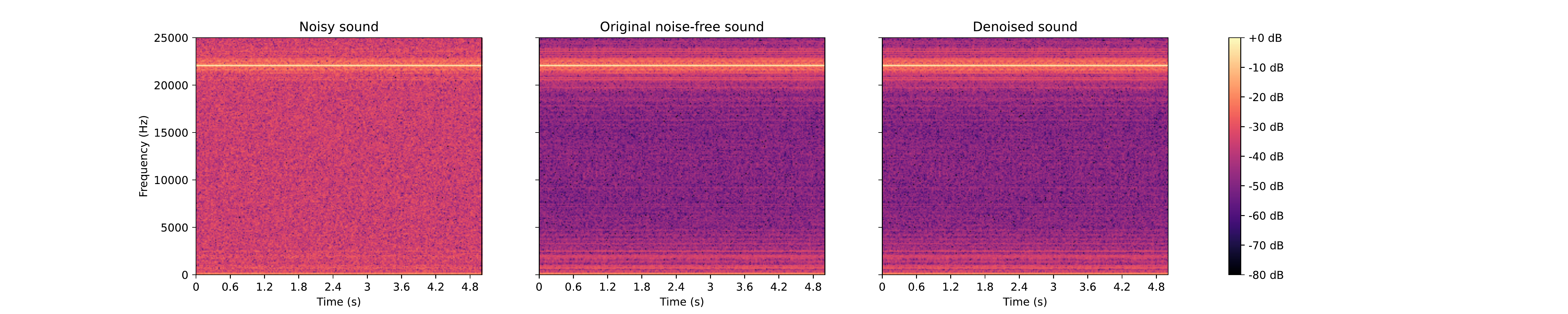}} 
    \caption{The waveform and STFT power spectrum of a clean sound in the test set, the corresponding noisy sound (random noise), and the denoised sound on the normal function category.}
    \label{fig:result_randomnoise}
\end{figure*}

\begin{figure*}[hpt]
    \subfigure[]{\includegraphics[width=0.9\textwidth]{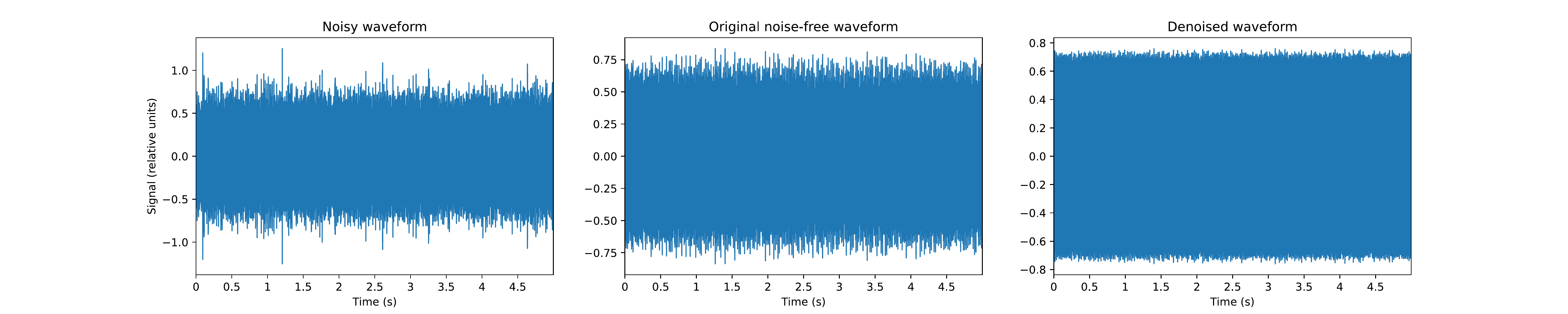}} 
    \subfigure[]{\includegraphics[width=\textwidth]{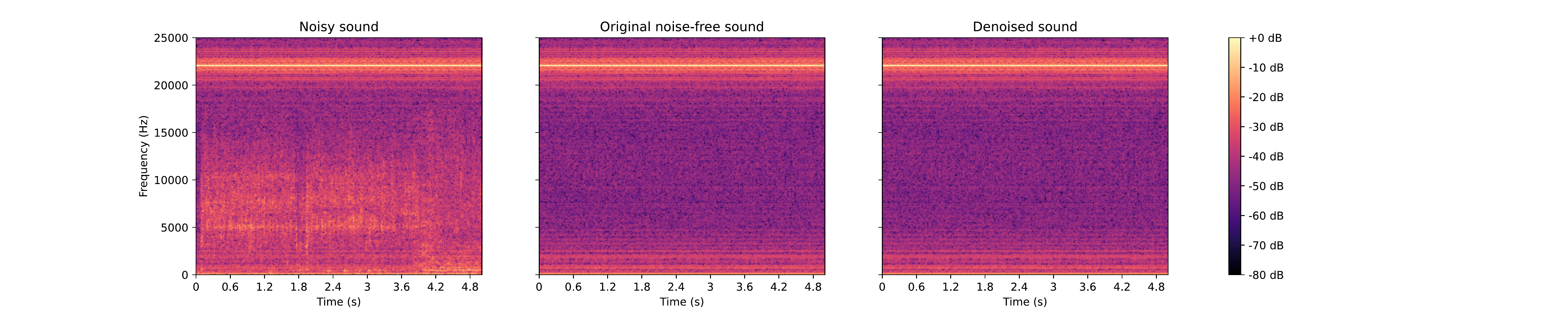}} 
    \caption{The waveform and STFT power spectrum of a clean sound in the test set, the corresponding noisy sound (the sound with water sink faucet noise), and the denoised sound on the normal function category.}
    \label{fig:result_sinkfaucet}
\end{figure*}

The MSE when denoising random Gaussian noise of the test set in the normal function category is in the range of 0.11 and 0.14, as shown in Fig.~\ref{fig:mse_normalCategory}a. The x-axis shows the 15 testing sounds in the test set and the y-axis shows the MSE values. 

\begin{figure*}[hpt]
    \centering
    \subfigure[]{\includegraphics[width=0.4\textwidth]{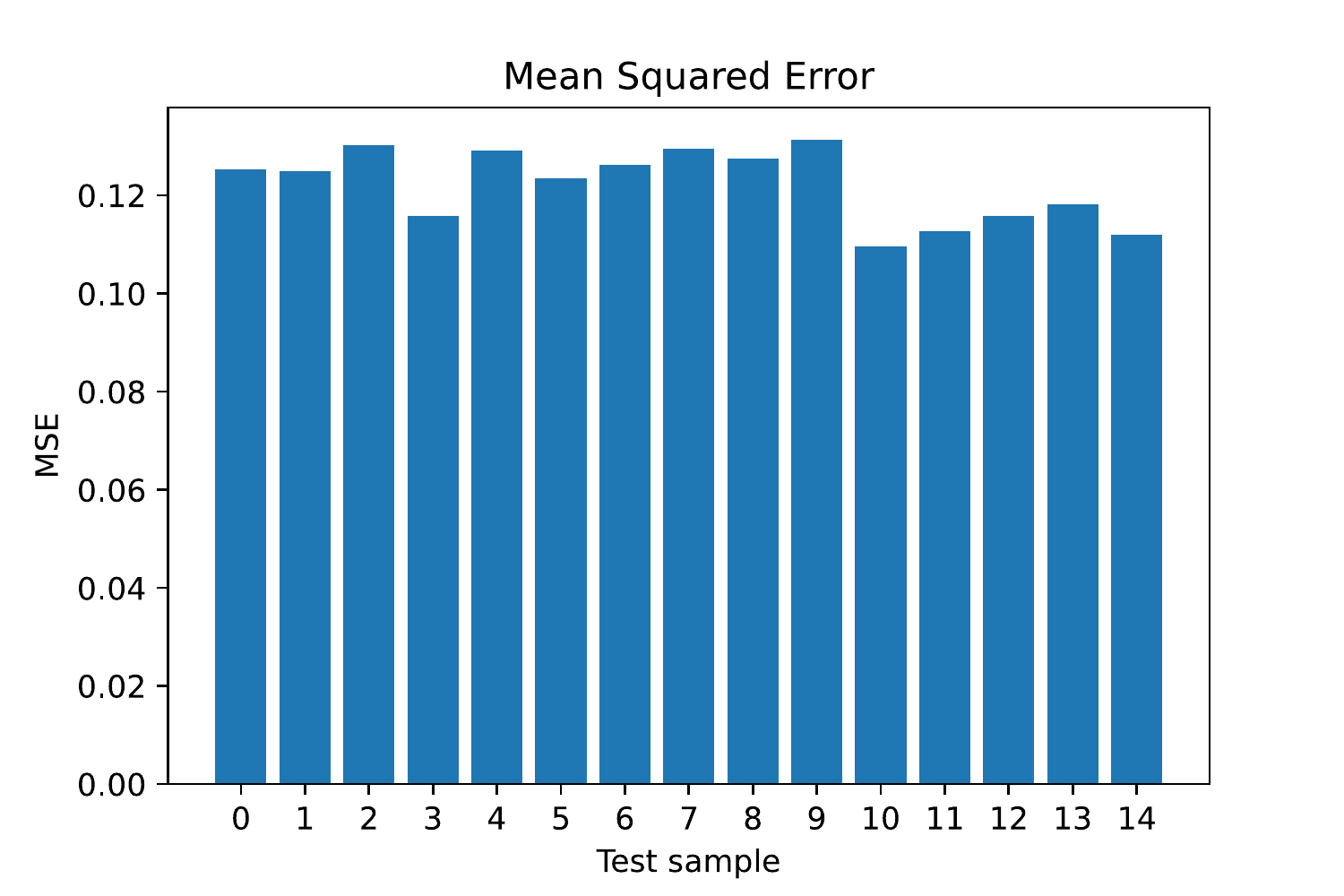}} 
    \subfigure[]{\includegraphics[width=0.4\textwidth]{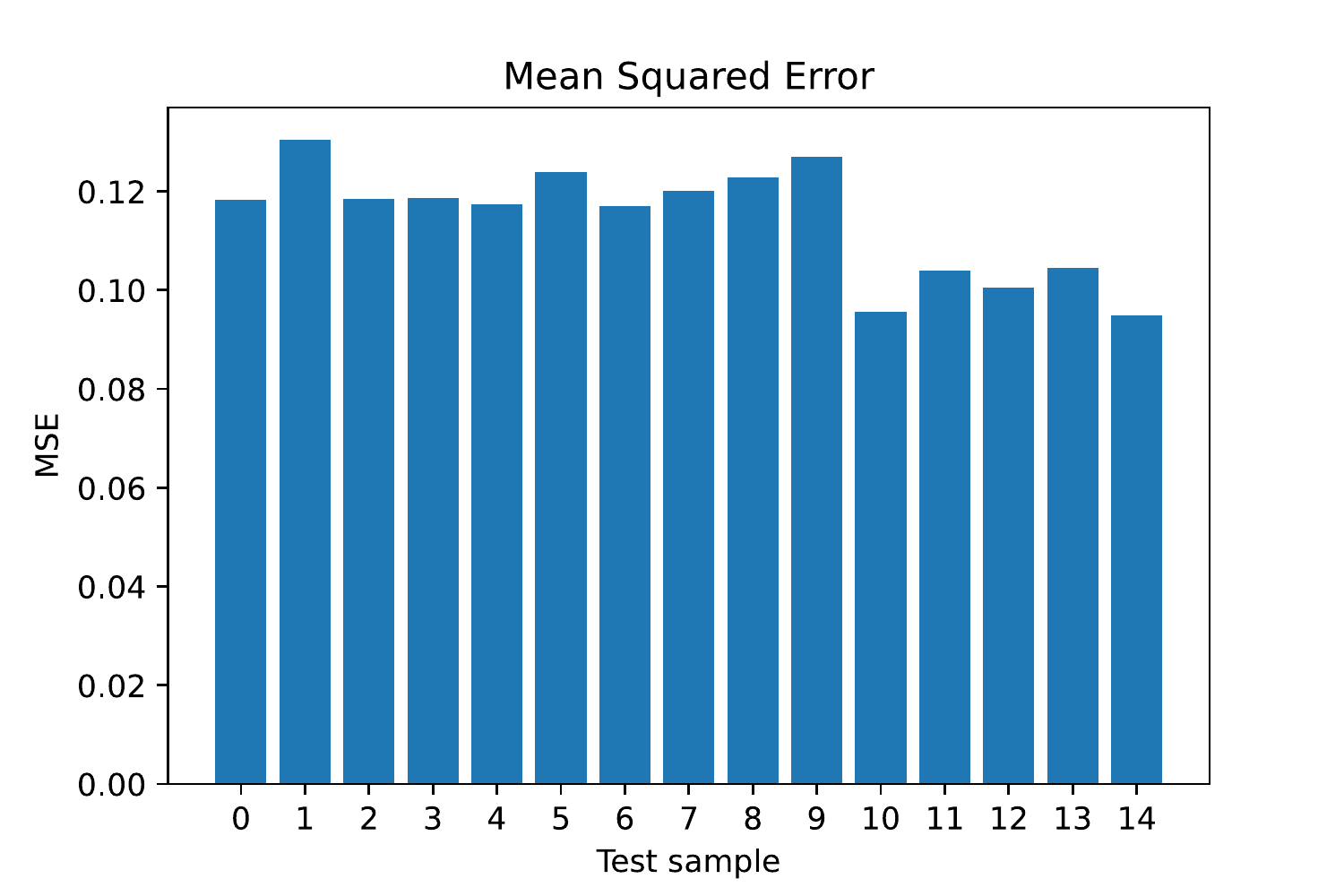}} 
    \caption{The MSE on the test set of the normal function category when removing: (a) the random Gaussian noise  (b) the water sink faucet noise.}
    \label{fig:mse_normalCategory}
\end{figure*}

\subsubsection{Denoise water sink faucet noise}
The waveform of the original noise-free sounds, noisy sounds, and denoised sounds are visualized in Figure~\ref{fig:result_sinkfaucet}a. The corresponding STFT power spectrum is visualized in Figure~\ref{fig:result_sinkfaucet}b. According to Fig. Figure~\ref{fig:result_sinkfaucet}b, the STFT power spectrum of the denoised sound using autoencoder is quite similar to the original noise-free sound. 
The MSE when denoising 15 testing sounds with water sink faucet noise in the normal function category is in the range of 0.09 and 0.14, as shown in Fig.~\ref{fig:mse_normalCategory}b.

\subsection{The horizontal misalignment fault category}
The same proposed autoencoder was trained and tested with the horizontal misalignment fault category. Seventy percent of the horizontal misalignment fault sounds were used for training and validation (137 sounds) and thirty percent for testing (60 sounds). The training and validation loss of the autoencoder when denoise random Gaussian noise and water sink faucet noise from sounds in the horizontal misalignment fault category is visualized in Figure~\ref{fig:loss_fault}a and Figure~\ref{fig:loss_fault}b, respectively. The autoencoder converges to a validation loss of 0.60 after 2 epochs.
\begin{figure*}[hpt!]
    \centering
    \subfigure[]{\includegraphics[width=0.4\linewidth]{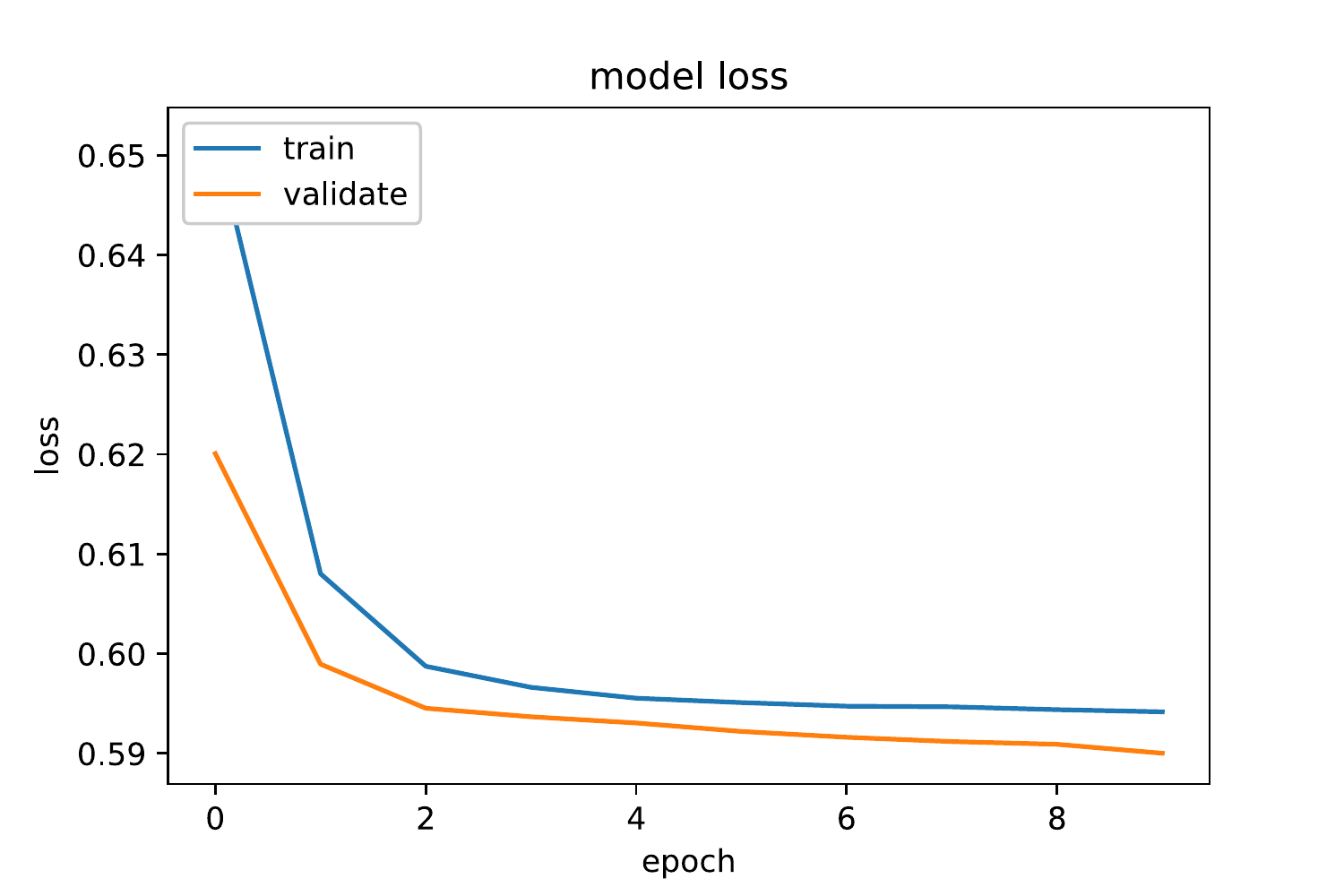}} 
    \subfigure[]{\includegraphics[width=0.4\linewidth]{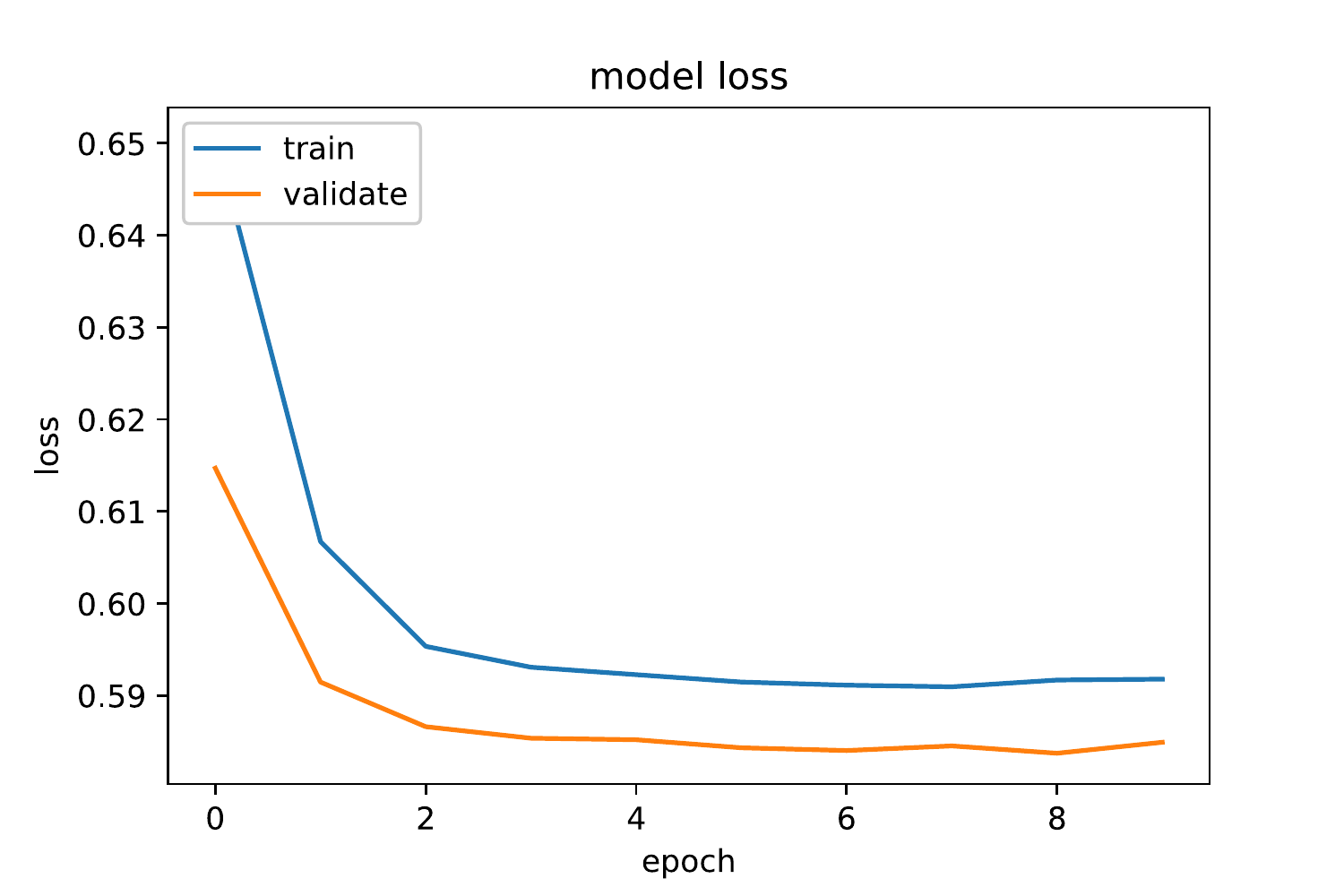}} 
    \caption{Training and validation loss for removing: (a) the random Gaussian noise (b) the water sink faucet noise on the horizontal misalignment fault category.}
    \label{fig:loss_fault}
\end{figure*}

\subsubsection{Denoise random Gaussian noise}
The waveform of the original noise-free sounds, noisy sounds, and denoised sounds are visualized in Figure~\ref{fig:result_randomnoise_fault}a. The corresponding STFT power spectrum is visualized in Figure~\ref{fig:result_randomnoise_fault}b.
\begin{figure*}[hpt!]
    \subfigure[]{\includegraphics[width=0.9\textwidth]{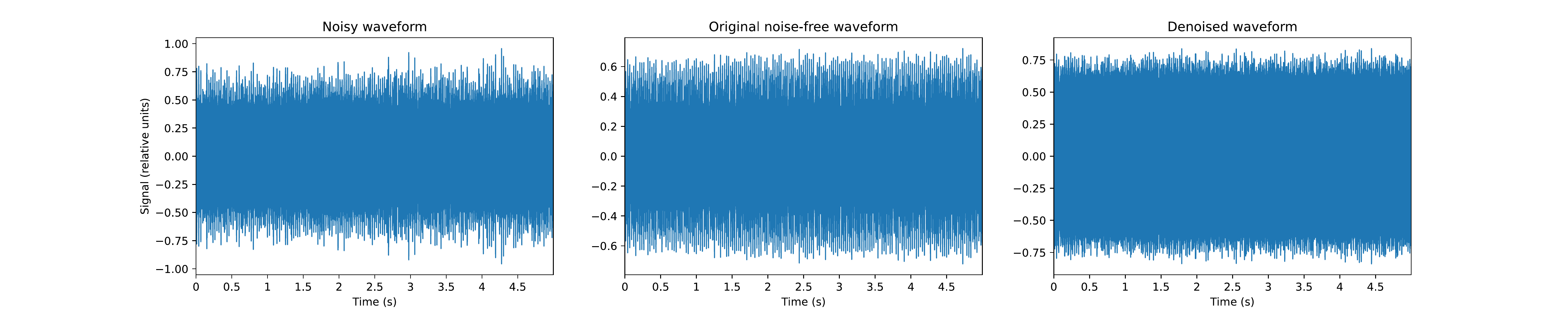}} 
    \subfigure[]{\includegraphics[width=\textwidth]{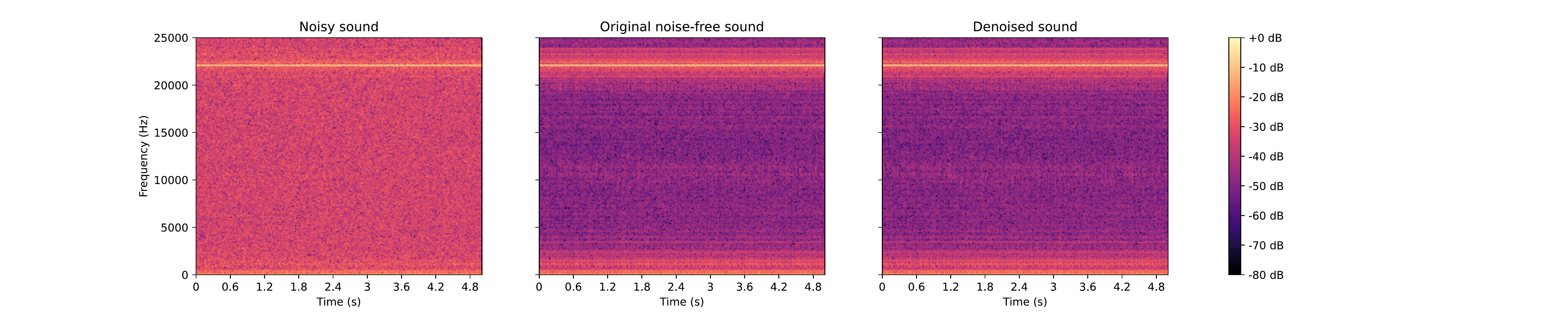}} 
    \caption{The waveform and STFT power spectrum of a clean sound in the test set, the corresponding noisy sound (random noise), and the denoised sound on the horizontal misalignment fault category.}
    \label{fig:result_randomnoise_fault}
\end{figure*}
The MSE of 60 horizontal misalignment fault sounds on the test set is in the range of 0.05 and 0.12, as shown in Fig.~\ref{fig:mse_waterSinkFaucet_fault}a.
\subsubsection{Denoise water sink faucet noise}

The waveform of the original noise-free sounds, noisy sounds, and denoised sounds are visualized in Figure~\ref{fig:result_sinkfaucet_fault}a. The corresponding STFT power spectrum is visualized in Figure~\ref{fig:result_sinkfaucet_fault}b.
\begin{figure*}[hpt!]
    \subfigure[]{\includegraphics[width=0.9\textwidth]{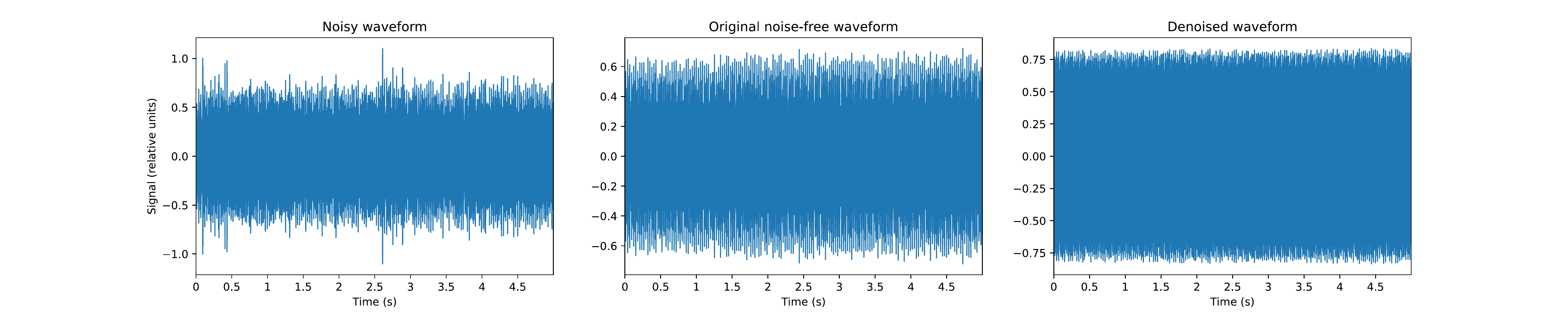}} 
    \subfigure[]{\includegraphics[width=\textwidth]{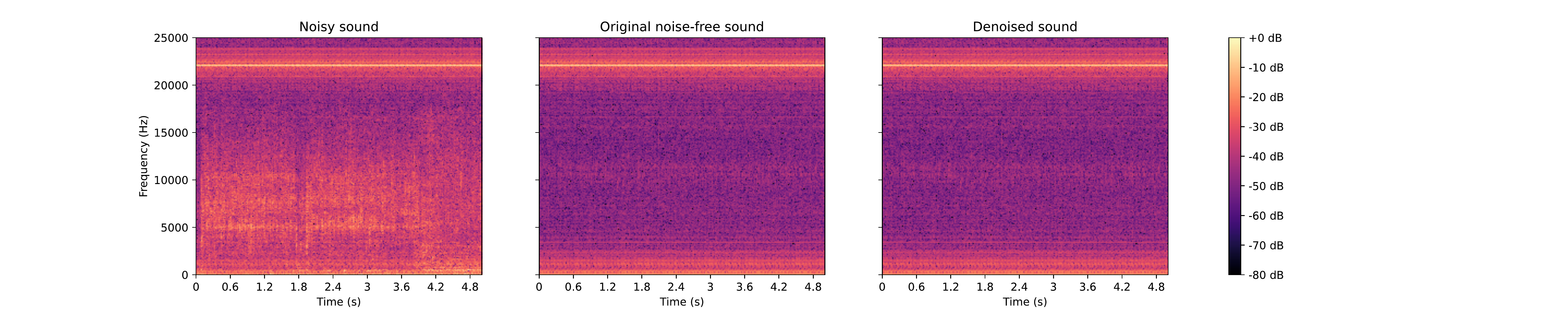}} 
    \caption{The waveform and STFT power spectrum of a clean sound in the test set, the corresponding noisy sound (the sound with water sink faucet noise), and the denoised sound on the horizontal misalignment fault category.}
    \label{fig:result_sinkfaucet_fault}
\end{figure*}
The MSE of 60 horizontal misalignment fault sounds on the test set is in the range of 0.09 and 0.15, as shown in Fig.~\ref{fig:mse_waterSinkFaucet_fault}b.

\begin{figure*}[hpt!]
    \centering
    \subfigure[]{\includegraphics[width=0.4\textwidth]{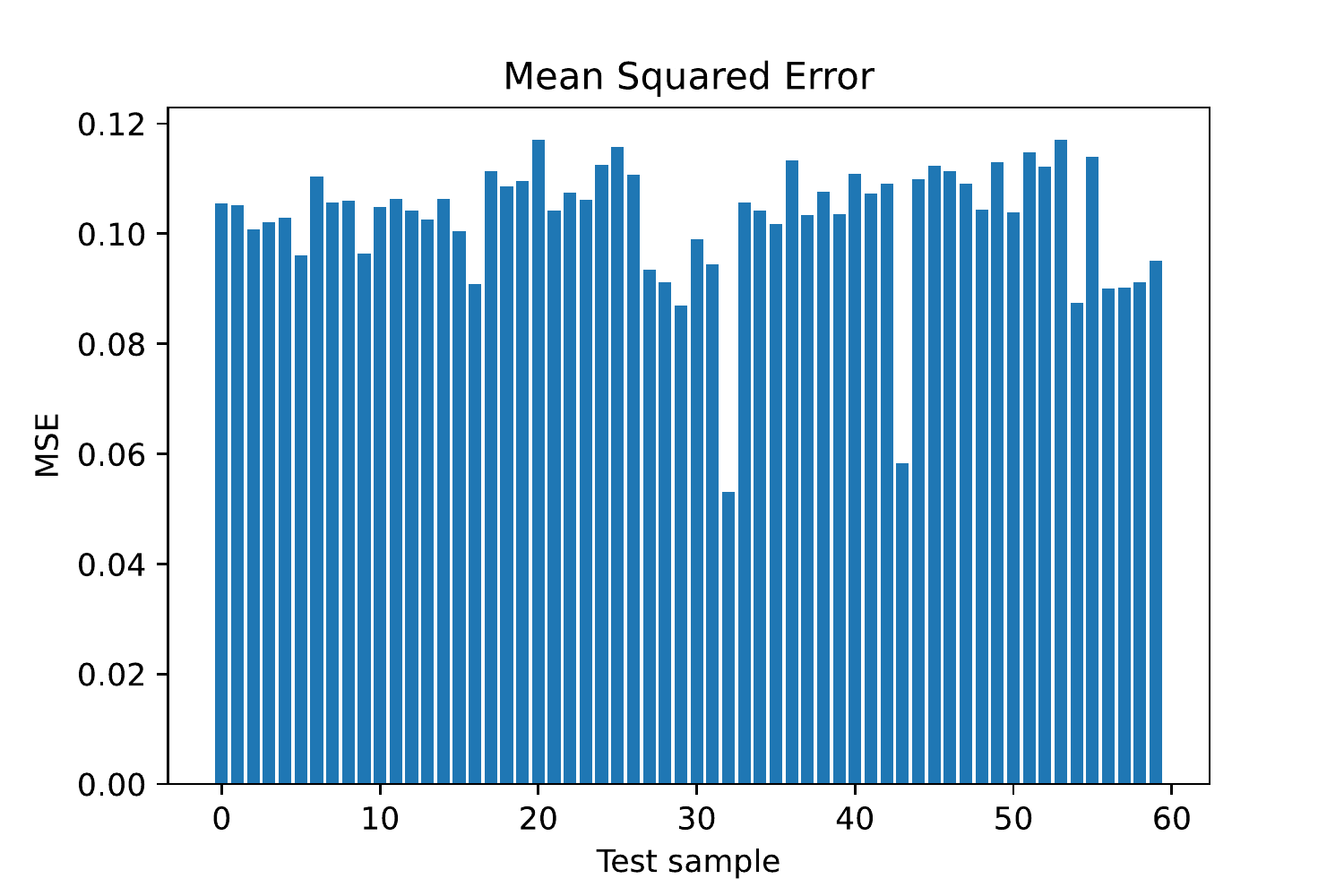}} 
    \subfigure[]{\includegraphics[width=0.4\textwidth]{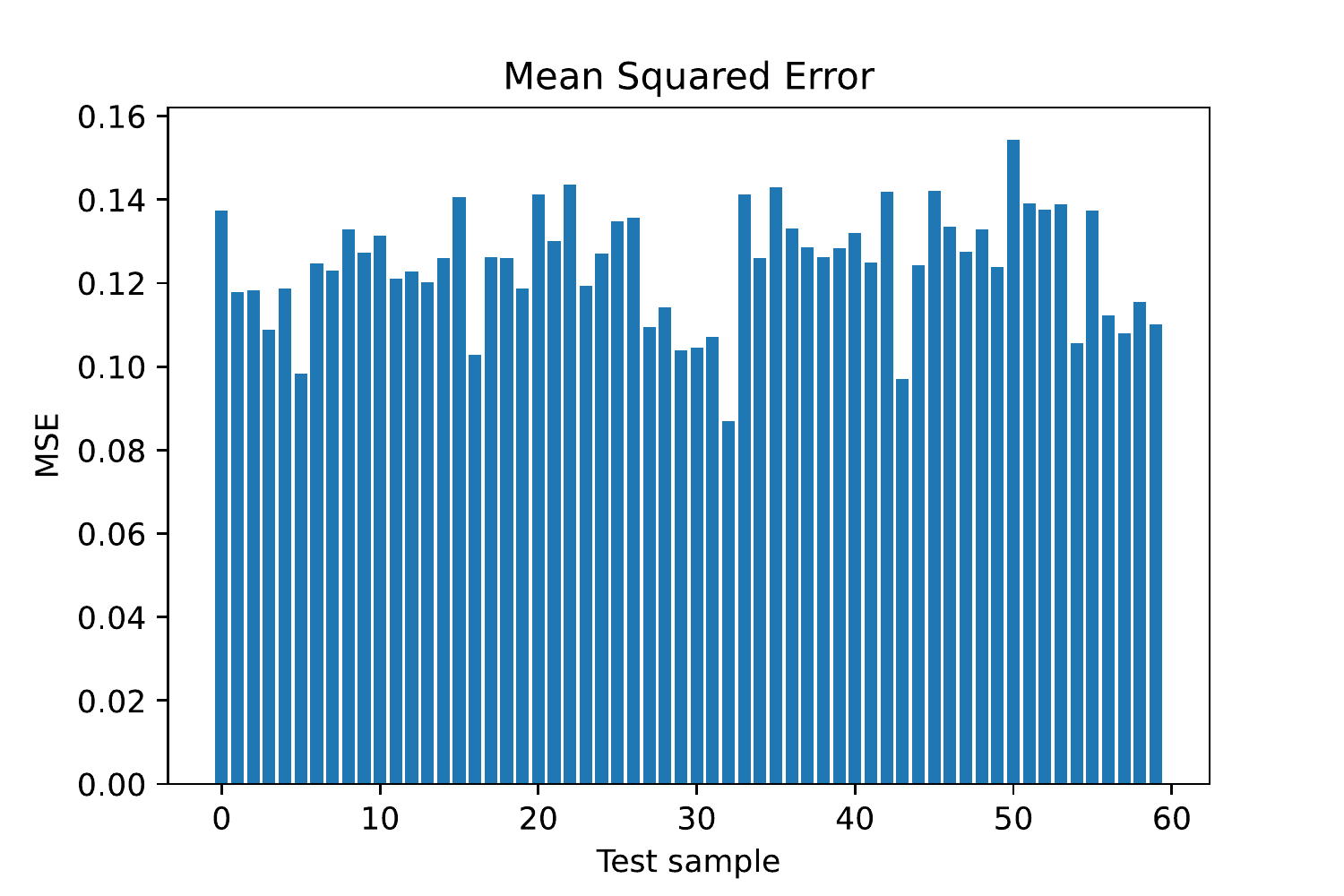}} 
    \caption{The MSE on the test set of the horizontal misalignment fault category when removing: (a) the random Gaussian noise  (b) the water sink faucet noise.}
    \label{fig:mse_waterSinkFaucet_fault}
\end{figure*}

\section{Discussion}

The low MSE indicates a high level of similarity between the denoised and original sounds. This means that the noise has been almost completely removed from the sound. The results were similar for all of the sounds in the test set.

When denoising sounds in the normal function category, the MSE on the test set demonstrates that the proposed autoencoder performed better with the water sink faucet noise than the random Gaussian noise (Figure 6). The reason for this is that the waveforms of normal sounds are relatively similar, and to train the denoising autoencoder, the same type of water sink faucet noise is added to all normal sounds. As a result, the trained autoencoder can denoise water sink faucet noises with ease. Meanwhile, because random Gaussian noise is generated at random, sounds using the random Gaussian noise may have a variety of waveforms.

In the horizontal misalignment fault category, the suggested autoencoder denoises the random Gaussian noise better than the water sink faucet noise (Figure 10). The reason for this is that there are four different sorts of horizontal misalignment fault sounds in the horizontal misalignment fault category (0.5mm, 1.0mm, 1.5mm, and 2.0mm). As a result, the waveform of sounds in this category differs, making autoencoder training to locate a common pattern amongst these corresponding noisy sounds more difficult.

The water sink faucet noise is clearly audible and overpowers the motor sound when we listen to the noisy sound with the water sink faucet noise. Despite the fact that the autoencoder was able to greatly minimize the flowing water sounds, some remained after the denoising process.




\section{Conclusion}
In this paper, we proposed an autoencoder to denoise sounds in order to improve machine fault diagnosis. Two types of noise (the random noise and the water sink faucet noise) were added to the original sounds to train a denoising autoencoder. The trained autoencoder can reconstruct noisy sounds that are almost the same as original noise-free sounds. This autoencoder generates an entirely new audio signal that has the background noise removed with the minimum of distortion. For the qualitative assessment, despite only displaying a few sounds (the waveform and corresponding STFT power spectrum) from the test sets in the results section, similar results were observed for all sounds in the test sets. For the quantitative assessment, the MSE for each sound in the test sets was assessed. The trained autoencoder denoised the water sink faucet noise better than the random Gaussian noise in the test set of the normal function category. Whereas, it denoised the random Gaussian noise better than the water sink faucet noise in the horizontal misalignment fault category.

This method is promising to reduce noise in sounds before processing it further in machine failure detection systems. For example, if we know a specific type of noise in the factory where the sound is recorded, we can add this specific noise into machine sounds and train an autoencoder to denoise this specific noise.

In the scope of this research, only 49 sounds in the normal function category and  197  horizontal misalignment fault sounds of the Machinery Fault Database have been used. Similarly, the proposed technique could be applied to the other classes in the Machinery Fault Database. While each model has been trained against a specific category of the machine sound dataset, this solution can also be improved by training a model to denoise all categories of the dataset. However, the autoencoder might not be able to reduce the noise in sounds perfectly. In some cases, the amount of loss tolerable in the reconstructed output becomes a trade-off decision. Such decisions are important in anomaly detection applications. The worst-case scenario would be if the autoencoder removes important features from an “abnormal” sound as noise, which then results in the sound being classified as a “normal” sound. During these instances, it is necessary to continually monitor the performance and update hyper-parameters of the networks so that the denoised autoencoder can reduce noise in sounds productively while still retaining distinguishing characteristics. These aspects need to be investigated further in continued studies.




\section*{Acknowledgment}
This research was supported by the Acoustic sensor set for AI monitoring systems (AISound ) project. The computations/data handling was enabled by resources provided by the Swedish National Infrastructure for Computing (SNIC) at the SNIC center is partially funded by the Swedish Research Council through grant agreement no. 2018-05973. We thank Thomas Svedberg at SNIC center for his assistance with concerning technical and implementational aspects, which were made possible through application support provided by SNIC.

\Urlmuskip=0mu plus 1mu\relax
\bibliographystyle{IEEEtran}
\bibliography{MyCollection.bib}
\end{document}